\documentclass[aps,pre,twocolumn,groupedaddress,showpacs]{revtex4}
\usepackage{graphics}
\usepackage{graphicx}
\usepackage{dcolumn}
\usepackage{bm}
\usepackage{amsfonts}
\begin{document}
\title{Nonlinear bending of molecular films by polarized light}
\author{Yu.~B.~Gaididei$^{1}$}
\author{A.~P.~Krekhov$^{2}$}
\email[]{alexei.krekhov@uni-bayreuth.de}
\author{H.~B{\"u}ttner$^{2}$}
\affiliation{$^{1}$Bogolyubov Institute for Theoretical Physics, 03143 Kiev, Ukraine}
\affiliation{$^{2}$Physikalisches Institut, Universit\"at Bayreuth, D-95440 Bayreuth, Germany}
\date{ \today }
\begin{abstract}
A theory of photoinduced directed bending of {\em non-crystalline} molecular 
films is presented.
Our approach is based on elastic deformation of the film due to interaction 
between molecules ordered through polarized light irradiation.
The shape of illuminated film is obtained in the frame of the nonlinear 
elasticity theory.
It is shown that the shape and the curvature of the film  depend on the 
polarization and intensity of the light.
The curvature of an irradiated film is a non-monotonic function of the 
extinction coefficient.
\end{abstract}
\maketitle
%
%
%
% Main text of the paper
%
\section{Introduction}
\label{intro}
Polymer films and solids containing light-sensitive molecules have the 
remarkable property to change their shape and size when irradiated with light.
Certain polymer films containing azobenzene chromophores in the main and side 
chains exhibit strong surface relief features under illumination: trenches under 
the action of linear polarized light and mounds or wells under the action of 
circularly polarized light \cite{Bian_1998,Kumar_1998}.
Circular azobenzene polyester films freely lying on a water surface become 
elliptically deformed under the influence of linearly polarized light 
\cite{Bublitz_2000}.
Large, reversible shape changes can be induced optically by photoisomerization of 
nematic elastomers \cite{Finkelmann_2001}.
Anisotropic bending and unbending behavior of molecular liquid-crystalline films  
containing azobenzene chromophores has been discovered and studied in 
Refs.~\cite{Ikeda_2003,Yu_2003} where it was shown that the films can be repeatedly 
and precisely bent along any chosen direction by using linearly polarized light.
Fast (on the timescale of $10^{-2}$~s) light induced bending of monodomain liquid 
crystal elastomers has been observed observed in \cite{Camacho_Lopez_2004}.
Shape-memory effects in polymers containing cinnamic groups induced by ultraviolet 
light illumination were reported quite recently in Ref.~\cite{Lendlein_2005}.
The possibility of coupling between orientational and translational degrees of 
freedom in liquid crystals was first raised by de Gennes \cite{DeGennes_1975} and 
extended to nematic elastomers in \cite{Warner_2003}.
Based on this idea a phenomenological theory of photoinduced deformations of 
nematic elastomers was proposed in \cite{Warner_2004}.
A microscopic theory of photoinduced deformation of non-crystalline molecular films 
was developed in \cite{Gaididei_2002}.
The physical reason for surface-relief formation presented in \cite{Gaididei_2002} 
is that azo-dyes have two isomeric states: {\em cis} and {\em trans}.
The molecules in these two states have significantly different shapes.
For example, in the case of azobenzene chromophores the {\em trans}-isomer is 
highly anisotropic whereas the {\em cis}-isomer is approximately isotropic 
\cite{Pedersen_1998}, so the multipole moments and sizes may differ significantly.
It was shown that there are two contributions to the photoelastic interaction: from 
the orientational interaction between molecules and from the interaction which is 
due to the change of the van der Waals interaction energy between a molecule and 
all surrounding molecules in its transition to the {\em cis}-isomer state.
The former causes the film deformation under the action of linearly polarized light 
while the latter (together with the orientational interaction) is responsible for 
the surface relief formation under the action of circularly polarized light.
The possibility of creation of wells and humps on the film surfaces under the 
action of circularly polarized light was discussed.
It is worth noting that in the frame of this approach neither orientational order 
nor orientational in-plane anisotropy of the film in the absence of irradiation was 
assumed.
One can say that the absorption of linearly polarized light creates an 
orientational order in the film which in turn produces anisotropic deformation.
Quite recently, based on the idea of isomeric states, a theory of the 
polarization-dependent photocontractions of polydomain elastomers due to 
light-induced director rotation, was proposed in \cite{Corbett_2006}.
The aim of the present report is to apply this approach to anisotropic bending of 
molecular films by polarized light.
We present an elastic energy of the film in the presence of polarized light and 
solving equations of equilibrium, we show that a change in the polarization 
direction of light causes a corresponding change of the shape of the film.
We also show that the curvature of an irradiated film is a non-monotonic function 
of the extinction coefficient.
\section{Elastic energy of molecular film}
\label{elastic}
We consider a film containing molecules with two different isomeric states.
Let the middle surface of the film coincide with the x,y-plane so that the 
undeformed film occupies the region: $|z| \leq \frac{h}{2}$, $x,y \in \Omega$.
The film is irradiated by a linearly polarized electromagnetic wave which 
propagates along the z-axis. 
Its electric component has the form
\begin{eqnarray}
\label{elw}
\vec{E}(\vec{r},t)=\vec{{\cal E}} \,\cos\left(k \,z-\,\omega\, t\right).
\end{eqnarray}
where $\omega$ is the frequency, $k=\omega/c$ is the wave number ($c$ is the speed 
of light) and
\begin{eqnarray}
\label{ore}
\vec{{\cal E}}={\cal E}\,\left(\cos \psi, \sin \psi, 0\right)
\end{eqnarray}
is the amplitude.
The angle $\psi$ determines the polarization of the electromagnetic wave.
The total elastic energy of irradiated thin film may be written as follows
\begin{eqnarray}
\label{eltot}
F = F_{el} + W_g + W \;.
\end{eqnarray}
Here the first term represent the elastic energy of the non-irradiated film
\begin{eqnarray}
\label{elast}
F_{el} = \frac{E}{2(1-\sigma^2)} \int\limits_{-h/2}^{h/2} dz 
\int\limits_{\Omega} dx dy
\Big[ \epsilon_{xx}^2 + \epsilon_{yy}^2
\nonumber\\
+ 2 \sigma \, \epsilon_{xx} \, \epsilon_{yy}
+ 2 (1-\sigma) \, \epsilon_{xy}^2 \Big] \;,
\end{eqnarray}
where $E$ and $\sigma$ are Young's modulus and Poisson's ratio, respectively 
\cite{Landau_1986} and $\epsilon_{\alpha\,\beta}$ is the strain tensor 
($\alpha,\beta=x,y$).
We assumed here that that the non-irradiated film is neither orientationally nor 
translationally ordered in the $xy$-plane and therefore we modeled its elastic 
properties by using the isotropic energy (\ref{elast}).
The term 
\begin{eqnarray}
\label{grav}
W_g = P \int\limits_{\Omega} dx dy \; w
\end{eqnarray}
in Eq.(\ref{eltot}) presents the potential energy of the film in the gravitational 
field.
Here $P=a_{gr}\,\rho_f\,h$ is the gravity force with $\rho_f$ being the film 
density and $a_{gr}$ being the acceleration of free fall.
The last term in Eq.(\ref{eltot}) gives the change of the elastic energy due to the 
interaction between the electromagnetic wave (\ref{elw}) and the film.
It has the form \cite{Gaididei_2002}
\begin{eqnarray}
\label{W}
W = W_1 + W_2 \;,
\end{eqnarray}
where
\begin{eqnarray}
\label{W1}
W_1 = - V_a \int\limits_{-h/2}^{h/2} dz 
\int\limits_{\Omega} dx dy \; {\cal N}(\vec{r}) 
\Big[ (\epsilon_{xx}-\epsilon_{yy}) \cos(2\psi)
\nonumber\\
+ 2 \epsilon_{xy} \sin(2\psi) \Big]
\end{eqnarray}
represents the anisotropic part of the photoelastic interaction which describes the 
coupling of the shear deformation of the film to the incoming light and
\begin{eqnarray}
\label{W2}
W_2 = - V_i \int\limits_{-h/2}^{h/2} dz 
\int\limits_{\Omega} dx dy \; {\cal N}(\vec{r})
(\epsilon_{xx}+\epsilon_{yy})
\end{eqnarray}
determines the isotropic in-plane deformations.
In Eqs.(\ref{W1})-(\ref{W2}) the parameter of the photoelastic interaction $V_a$ is 
due to the orientational (e.g., dipole-dipole) part of intermolecular interaction 
while the parameter $V_i$ is due to isotropic part of the intermolecular 
interaction.
For the sake of simplicity we assume them space independent.
The function ${\cal N}(\vec{r})$ gives the population of {\em cis}-isomers for a 
given value of the radiation power ${\cal E}^2$ \cite{Gaididei_2002}.
Following the Bouguer-Lambert-Beer law which determines how the intensity of light 
decreases under its propagation inside an absorbing medium, we shall model the 
function 
${\cal N}(\vec{r})$ as follows
\begin{eqnarray}
\label{popul}
{\cal N}(\vec{r}) = {\cal N}_0 \exp\left\{ \frac{(z - h/2)}{\xi} \right\} \;,
\end{eqnarray}
where ${\cal N}_0$ is the maximum population of the {\em cis}-isomers for a given 
power ${\cal E}^2~$ and $\xi$ is the extinction length of the light which provides 
transition of chromophores from {\em trans}- to {\em cis}-isomeric state (in the 
experiments \cite{Ikeda_2003,Yu_2003} it was the light with the wavelength 366 nm); 
other distribution can of course be used, depending on the actual arrangement of 
azo-dyes.
Note that in Eq.(\ref{popul}) we neglected the fact that upon bending the normal to 
the film surface deviates from the $z$-direction which is legitimate when bending 
is small.
Note that in the case of circularly polarized light when instead of 
Eqs.(\ref{elw}), (\ref{ore}) we have
\begin{eqnarray}
\label{elcirc} 
\vec{E}(\vec{r},t) = {\cal E} \,\left\{ \cos(k z - \omega t) \, , \,
\sin(k z - \omega t) \, , \, 0 \right\}
\end{eqnarray}
and therefore the contribution (\ref{W1}) vanishes; the energy of photoelastic 
interaction is solely determined by Eq.(\ref{W2}).
Following the usual derivation of F{\"o}ppl--von Karman equations for bending of a 
thin plate (see, e.g., \cite{Stoker_1983}), we write the strains as a linear 
expansion in $z$ from the middle plane and get
\begin{eqnarray}
\label{strain_exp}
\left(\begin{array}{lcr}
\epsilon_{xx}& \epsilon_{xy}\\ \epsilon_
{xy}& \epsilon_{yy}
\end{array}\right)=\left(\begin{array}{lcr}
u_{xx}& u_{xy}\\ u_{xy}& u_{yy}
\end{array}\right)
+ z \left(\begin{array}{lcr}
\partial_{xx}\,w& \; \partial_{xy}\,w\\ \partial_ {xy}\,w& \; \partial_{yy}\,w
\end{array}\right) \;,
\end{eqnarray}
where
\begin{eqnarray} 
\label{nonlstr}
u_{\alpha\beta}=\frac{1}{2}\,\left(\partial_{\beta}
u_{\alpha}+\partial_{\alpha}
u_{\beta}\right)+\frac{1}{2}\,\partial_{\alpha}
w\,\partial_{\beta} w
\end{eqnarray}
are the components of the two-dimensional nonlinear deformation tensor and
$\partial_\alpha$ denotes differentiation with respect to the coordinate
$x_\alpha = x, y$.
Referred to these coordinates, the components of displacement are
$u_\alpha = u_x, u_y, w$ where we have named the vertical displacement $u_z=w$.
Introducing this equations into Eq.(\ref{eltot}) we can represent the elastic 
energy of an irradiated film in the form
\begin{eqnarray}
\label{elt} 
F = F_b + F_s + W \;.
\end{eqnarray}
The bending energy is written as
\begin{eqnarray}
\label{elbend}
F_b = \frac{D}{2} \int\limits_{\Omega} dx dy \Big\{ \left( \Delta w \right)^2
\nonumber\\
+ 2 (1-\sigma) \left[ \left( \partial_{x y} w \right)^2
- \partial_x^2 w \, \partial_y^2 w \right] \Big\} \;,
\end{eqnarray}
where $\Delta=\partial_\alpha\,\partial_\alpha$ is the two-dimensional Laplace 
operator and
\begin{eqnarray}
D=\frac{E\,h^3}{12\,(1-\sigma^2)}
\end{eqnarray}
is the flexural rigidity of the film.
For the stretching energy one has
\begin{eqnarray}
\label{elstr} 
F_s = \frac{h}{2} \int\limits_{\Omega} dx dy
\left( \sigma_{x x} \, u_{x x} + \sigma_{y y} \, u_{y y}
+ 2 \sigma_{x y} \, u_{x y} \right)
\end{eqnarray}
with the longitudinal stresses
\begin{eqnarray}
\label{stress}
&& \sigma_{xx}=\frac{E}{1-\sigma^2} \left( u_{xx} + \sigma \, u_{yy} \right) \;,
\nonumber\\
&& \sigma_{yy}=\frac{E}{1-\sigma^2} \left( u_{yy} + \sigma \, u_{xx} \right) \;,
\nonumber\\
&& \sigma_{xy}=\frac{E}{1+\sigma} \, u_{xy} \;.
\end{eqnarray}
In Eq.(\ref{elt}) the light-film interaction energy becomes
\begin{eqnarray}
\label{totalint}
W = W_b + W_s
\end{eqnarray}
with the bending contribution
\begin{eqnarray}
\label{bendint}
W_b = W_{b 1} + W_{b 2} \;,
\end{eqnarray}
where
\begin{eqnarray}
\label{bendint1}
W_{b 1} = -h^2 \int\limits_{\Omega} dx dy
\Big\{ A_i \Delta w
\nonumber\\
+ A_a \Big[ \left(
\partial_x^2 w - \partial_y^2 w \right) \cos(2\psi) 
+ 2 \, \partial_{x y} w \, \sin(2\psi) \Big] \Big\}
\end{eqnarray}
describes a linear interaction with bending deformation while the term
\begin{eqnarray}
\label{bendint2}
W_{b 2} = -\frac{h}{2} \int\limits_{\Omega} dx dy 
\Big\{ B_a \Big[ \big( \left( \partial_x w \right)^2 
- \left( \partial_y w \right)^2 \big) \cos(2\psi)
\nonumber\\
+ 2 \, \partial_x w \, \partial_y w \, \sin(2\psi) \Big] 
+ B_i \big[ \left( \partial_x w \right)^2 + \left( \partial_y w \right)^2 \big]
\Big\} \;\;
\end{eqnarray}
is due to the nonlinear character of the two-dimensional deformation tensor 
(\ref{nonlstr}).
The stretching contribution in Eq.(\ref{totalint}) is
\begin{eqnarray}
\label{stretchint}
W_s = -h \int\limits_{\Omega} dx dy \Big\{ B_a 
\Big[(u_{xx}-u_{yy}) \cos(2\psi)
\nonumber\\
+ 2 \, u_{xy} \sin(2\psi) \Big] + B_i (u_{xx}+u_{yy}) \Big\} \;,
\end{eqnarray}
and the new light-film interaction parameters in 
Eqs.(\ref{bendint1})-(\ref{stretchint}) are
\begin{eqnarray}
\label{AB}
&& A_i=V_i\,\overline{z\,{\cal N}},\nonumber\\
&& A_a=V_a\,\overline{z\,{\cal N}},\nonumber\\
&& B_i=V_i\,\overline{{\cal N}},\nonumber\\
&& B_a=V_a\,\overline{{\cal N}}.
\end{eqnarray}
The function
\begin{eqnarray}
\label{mean}
\overline{{\cal N}} \equiv \frac{1}{h} \int\limits_{-h/2}^{h/2} 
{\cal N}(z) \, dz = {\cal N}_0 \, \frac{\xi}{h} \left( 1 - e^{- h/\xi} \right)
\end{eqnarray}
gives the mean value of {\em cis}-isomers in the film and
\begin{eqnarray}
\label{f}
\overline{{z \, \cal N}} \equiv \frac{1}{h^2} \int\limits_{-h/2}^{h/2} 
z {\cal N}(z) \, dz
\nonumber\\
=
\frac{1}{2} {\cal N}_0 \, \frac{\xi}{h} \, \Big[ 1 - 2 \frac{\xi}{h} + 
\left( 1 + 2 \frac{\xi}{h} \right) e^{- h/\xi} \Big]
\end{eqnarray}
characterizes asymmetry of the {\em cis}-isomer distribution in the film.
By using Green's formula for the two-dimensional integrals, Eqs.(\ref{bendint1}) 
and (\ref{stretchint}) can be presented in an equivalent form
\begin{eqnarray}
\label{contbend}
W_{b 1} = -h^2 \oint dl
\Big\{ \left[ A_i + A_a \cos{2(\psi-\theta)} \right] \frac{\partial w}{\partial n}
\nonumber\\
+ A_a \sin{2(\psi-\theta)} \frac{\partial w}{\partial l} \Big\} \;,
\end{eqnarray}
\begin{eqnarray}
\label{contstr}
W_s = -h \oint dl
\Big\{ B_i \, \vec{n}\cdot\vec{u} + B_a \cos(2\psi-\theta) u_x
\nonumber\\
+ B_a \sin(2\psi-\theta) u_y \Big\} \;,
\end{eqnarray}
where $\partial/\partial l$ is the derivative along the tangent $\vec{l}$ to the 
contour and it has together with the normal derivative $\partial/\partial n$ the 
form
\begin{eqnarray}
\label{norm}
&& \frac{\partial}{\partial l} = \cos\theta \, \partial_y - \sin\theta \, 
\partial_x \;,
\nonumber\\
&& \frac{\partial}{\partial n} = \cos\theta \, \partial_x + \sin\theta \, 
\partial_y \;.
\end{eqnarray}
where the angle $\theta$ determines the direction of the outward normal to the 
boundary contour: $\vec{n}=\left(\cos\theta,\sin\theta\right)$.
Eqs.(\ref{bendint1}) and (\ref{stretchint}) present two physically different 
mechanisms of film deflection.
Eq.(\ref{stretchint}) describes the light-film interaction which causes the change 
of the film area.
The intensity of the interaction is proportional to the mean value of 
{\em cis}-isomers in the film $\overline{{\cal N}}$ and as it is seen from 
Eq.(\ref{contstr}), the action of light is equivalent to a uniformly distributed 
edge force applied in the plane of the film.
In the presence of the interaction (\ref{stretchint}) with $ B_i > 0$ the area of 
the film increases while the opposite sign corresponds to the compression of the 
film.
On the other hand the interaction given by Eq.(\ref{bendint1}) is due to asymmetric 
distribution of {\em cis}-isomers in the film (\ref{f}).
As Eq.(\ref{contbend}) shows, in this case the light produces a bending moment 
applied to the boundary contour.
\section{Deflections under the action of polarized light}
\label{deflections}
The Euler-Lagrange equations for the functional (\ref{elt}), (\ref{elbend}), 
(\ref{elstr}), (\ref{totalint})-(\ref{stretchint}) (F{\"o}ppl-von-Karman equations) 
have the form
\begin{eqnarray}
\label{fvkl}
D\,\Delta^2\,w
-h\,\frac{\partial}{\partial\,x_{\beta}}\,\sigma_{\alpha\beta}\,
\frac{\partial \,w}{\partial\,x_{\alpha}}
+h\,B_i\,\Delta\,w
\nonumber\\
+h\,B_a\left[(\partial^2_x w-\partial^2_y w) \cos 2 \psi
+2 \, \partial_{x y} w \, \sin 2 \psi \right]
\nonumber\\
= -P \;,
\\
\label{fvk2}
\frac{\partial}{\partial\,x_{\beta}}\,\sigma_{\alpha\beta} = 0 \;.
\end{eqnarray}
The boundary conditions for these equations may be obtained in the same way as it 
was done in \cite{Landau_1986} and in the case of free boundary (the edge of the 
film is free) the variations of the vertical component $\delta w$ and its normal 
derivative $\partial (\delta w)/\partial n$ on the edge are arbitrary.
This gives the following set of equations
\begin{eqnarray}
\label{bcdwn}
D \Big[ \frac{\partial}{\partial n} \Delta w
+(1-\sigma) \frac{\partial}{\partial l} \{ \sin\theta\cos\theta
\left( \partial_y^2 w - \partial_x^2 w \right)
\nonumber\\
+\cos2\theta \partial_{xy} w \} \Big]
-h^2 A_a \frac{\partial}{\partial l} \sin2(\psi-\theta) = 0 \;,
\end{eqnarray}
\begin{eqnarray}
\label{bcdw}
D \Big[ \Delta w
-(1-\sigma) \{ \sin^2\theta \partial_x^2 w
+\cos^2\theta \partial_y^2 w
\nonumber\\
-\sin2\theta \partial_{xy} w \} \Big]
-h^2 \left[ A_i + A_a \cos2(\psi-\theta) \right]=0 \;,
\end{eqnarray}
\begin{eqnarray}
\label{bcdu}
\sigma_{x\beta}\,n_{\beta}\,-\,\left(B_i+B_a\,\cos2\psi\right)\,\cos\theta\,
\nonumber\\
-B_a\,\sin2\psi\,\sin\theta = 0 \;,
\nonumber\\
\sigma_{y\beta}\,n_{\beta}\,-\,\left(B_i-B_a\,\cos2\psi\right)\,\sin\theta\,
\nonumber\\
-B_a\,\sin 2\psi\,\cos\theta = 0 \;.
\end{eqnarray}
By introducing the Airy potential $\chi(x,y)$, so that Eqs.(\ref{fvk2}) are 
automatically satisfied:
\begin{eqnarray}
\label{airy}
\sigma_{xx}=\partial^2_y\,\chi \;,\;
\sigma_{yy}=\partial^2_x\,\chi \;,\;
\sigma_{xy}=\sigma_{yx}=-\partial_{xy}\,\chi \;,
\end{eqnarray}
and presenting the Airy potential as a sum
\begin{eqnarray}
\label{T}
\chi(x,y)=\frac{1}{2}\,(B_i - B_a\,\cos2\psi)\,x^2
\nonumber\\
+\frac{1}{2}\,(B_i + B_a\,\cos2\psi)\,y^2
\nonumber\\
-B_a\,\sin2\psi \, x y + T(x,y) \;,
\end{eqnarray}
we obtain the F{\"o}ppl-von-Karman equation (\ref{fvkl}) in the form
\begin{eqnarray}
\label{fvk}
D \Delta^2 w
-h \left( \partial^2_x T \, \partial^2_y w + \partial^2_y T \, \partial^2_x w
-2 \partial_{xy} T \, \partial_{xy} w \right)
\nonumber\\
= -P \;.
\end{eqnarray}
This equation is to be completed by the compatibility condition \cite{Landau_1986}
\begin{eqnarray}
\label{cc}
\Delta^2\,T
+\,E\,\left[ \partial^2_x\,w\,\partial_y^2\,w-\left(\partial_{xy}\,w\right)^2 
\right]
=0 \;.
\end{eqnarray}
Note that owing to the fact that we extracted from the Airy potential $\chi(x,y)$ 
the parabolic contribution, the F{\"o}ppl-von-Karman equation (\ref{fvk}) does not 
contain light-induced driving terms.
Introducing Eqs.(\ref{airy}), (\ref{T}) into Eqs.(\ref{bcdu}), we obtain that the 
boundary conditions for the potential $T(x,y)$ have particularly simple form:
\begin{eqnarray}
\label{bcT}
\partial^2_x\,T\,\sin\theta-\partial_{xy}T\,\cos\theta=0 \;,
\nonumber\\
\partial^2_y\,T\,\cos\theta-\partial_{xy}\,T\,\sin\theta = 0 \;.
\end{eqnarray}
\section{Bending of circular film}
\label{bend_circular}
We consider the photoinduced bending of a circular film of radius R.
In the polar coordinates 
\begin{eqnarray}
x=\rho\,R\,\cos\theta \;,\;\; y=\rho\,R\,\sin\theta \;,
\end{eqnarray}
where $\rho=r/R$, ($r=\sqrt{x^2+y^2}$) is a dimensionless radial coordinate and 
$\theta$ is the azimuthal angle we obtain from Eqs.(\ref{fvk}) and (\ref{cc}) that 
equations for new dimensionless dependent variables
\begin{eqnarray}
\label{newwt}
\zeta=\frac{w}{h} \;, \;\;
\tau=\frac{h}{D} T
\end{eqnarray}
take the following form
\begin{eqnarray}
\label{fvkpol}
\Delta^2 \zeta
-\frac{1}{\rho^2} \Big\{ \left( \partial_\theta^2 \zeta
+\rho \partial_\rho \zeta \right) \partial^2_\rho \tau
+\left(\partial_\theta^2 \tau + \rho \partial_\rho \tau \right) \partial^2_\rho 
\zeta
\nonumber\\
-\frac{2}{\rho^2} \left( \partial_\theta \tau
-\rho \partial_{\rho\theta} \tau \right) \left( \partial_\theta \zeta
-\rho \partial_{\rho\theta} \zeta \right) \Big\}
\nonumber\\
=-f \;,
\end{eqnarray}
\begin{eqnarray}
\label{ccpol}
\Delta^2\,\tau
+ 12(1-\sigma^2)\frac{1}{\rho^2}\,\Big\{\left(\partial_\theta^2 \zeta
+\rho\,\partial_\rho\zeta\right)\,\partial^2_\rho\zeta
\nonumber\\
-\frac{1}{\rho^2}\,\left(\partial_\theta \zeta
-\rho\,\partial_{\rho\theta}\zeta\right)^2\Big\}
=0 \;.
\end{eqnarray}
The boundary conditions (\ref{bcdwn}), (\ref{bcdw}) and (\ref{bcT}) at $\rho=1$ 
($r=R$) become
\begin{eqnarray}
\label{bcdwnpol}
\,\Big[\partial_\rho\Delta\,\zeta
+\frac{1-\sigma}{\rho}\partial_\rho
\left(\frac{1}{\rho}\partial^2_\theta \zeta \right) \Big]
\nonumber\\
-\frac{2}{\rho} a_a \,
\cos2(\psi-\theta)
=0 \;,
\end{eqnarray}
\begin{eqnarray}
\label{bcdwpol}
\,\Big[\partial^2_\rho \zeta
+\sigma\,\left(\frac{1}{\rho}\,\partial_\rho\zeta
+\frac{1}{\rho^2}\,\partial^2_\theta \zeta\right)\Big]
\nonumber\\
-\,\left[a_i + a_a\,\cos2(\psi-\theta)\right]
=0 \;,
\end{eqnarray}
\begin{eqnarray}
\label{bcTpol}
(1-\rho\partial_\rho)\,\partial_\theta\,\tau = 0 \;,
\nonumber\\
\rho\,\partial_\rho\,\tau+\partial^2_\theta \,\tau = 0 \;,
\end{eqnarray}
where $f=P R^4/(h D)$ is a dimensionless gravity force and dimensionless parameters
\begin{eqnarray}
\label{ai_aa}
a_i = A_i \frac{h R^2}{D} \;, \;\;
a_a = A_a \frac{h R^2}{D} \;
\end{eqnarray}
characterize the intensity of the light-film interaction.
To have some insight we consider first the case of weak light-film interaction: 
$a_i<1$, $a_a<1$.
Then we compare these results with numerics.
\subsection{Circularly polarized light}
\label{circular}
\begin{figure}
\resizebox{0.75\columnwidth}{!}{
\includegraphics{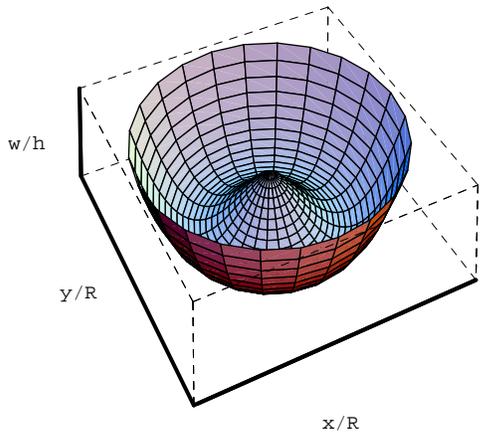}
}
\caption{The shape of the circular film illuminated by a circularly polarized 
light.}
\label{fig:1}
\end{figure}

In this subsection we consider the case of photoinduced film deformation when the 
light is circularly polarized (\ref{elcirc}) (the parameter $a_a$ is set to zero).
Assuming the radial symmetry of solutions 
($\partial_\theta\zeta=\partial_\theta\tau=0$) from 
Eqs.(\ref{fvkpol})-(\ref{bcTpol}) one can obtain approximately (see Appendix for 
details) that the shape of an irradiated film is determined by the expression
\begin{eqnarray}
\label{wrad}
\zeta_0(\rho) = \frac{\rho^2}{2}\Big\{\frac{a_i}{1+\sigma}
-\left(\frac{3+\sigma}{2\,(1+\sigma)}+\frac{\rho^2}{4}-2\ln\,\rho\right)\,
\frac{f}{8}\Big\} \;\;
\end{eqnarray}
(see Fig.~\ref{fig:1}).
%

%%% Figure 1 %%%
\begin{figure}
\resizebox{0.75\columnwidth}{!}{
\includegraphics{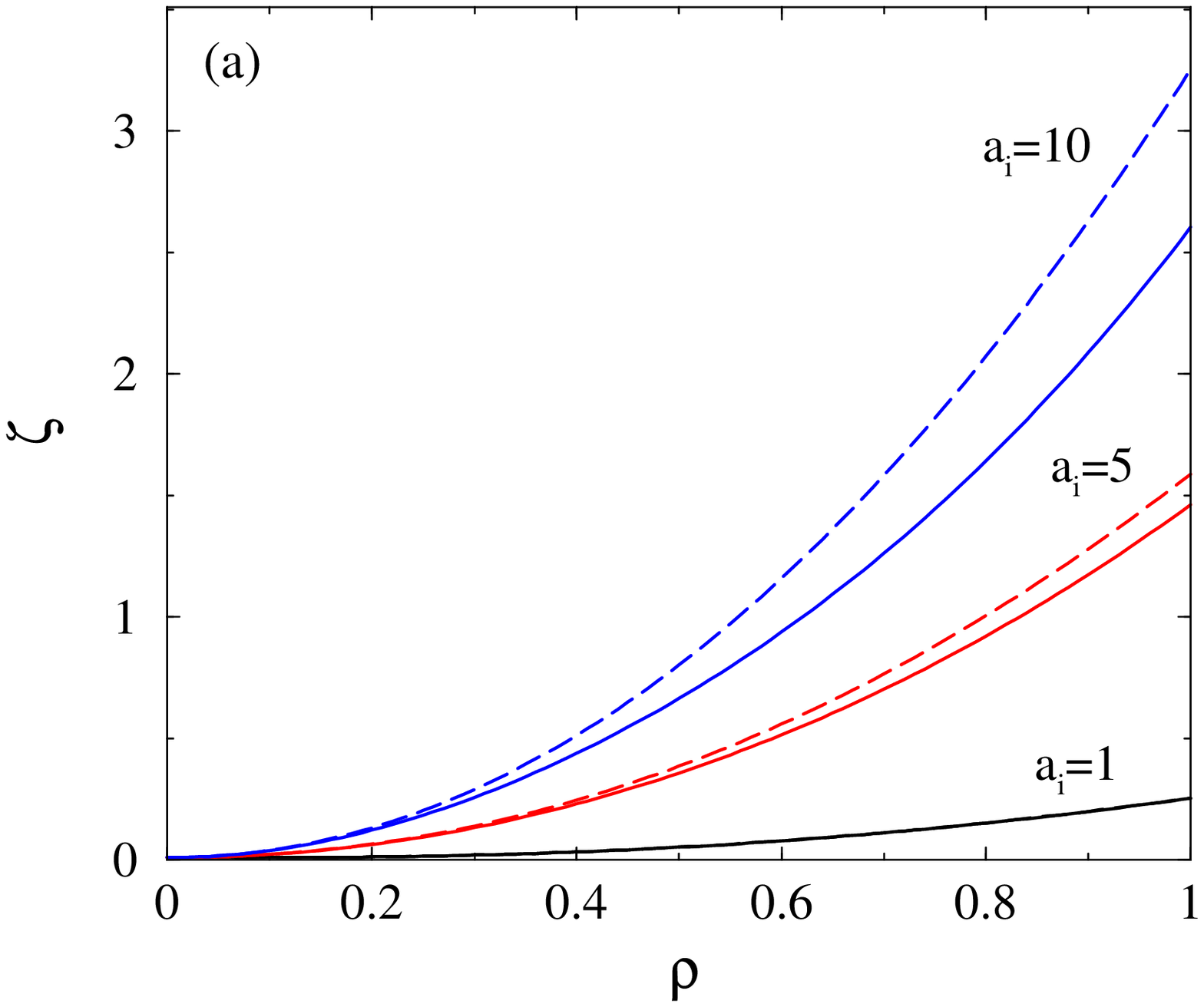}
}
\resizebox{0.75\columnwidth}{!}{
\includegraphics{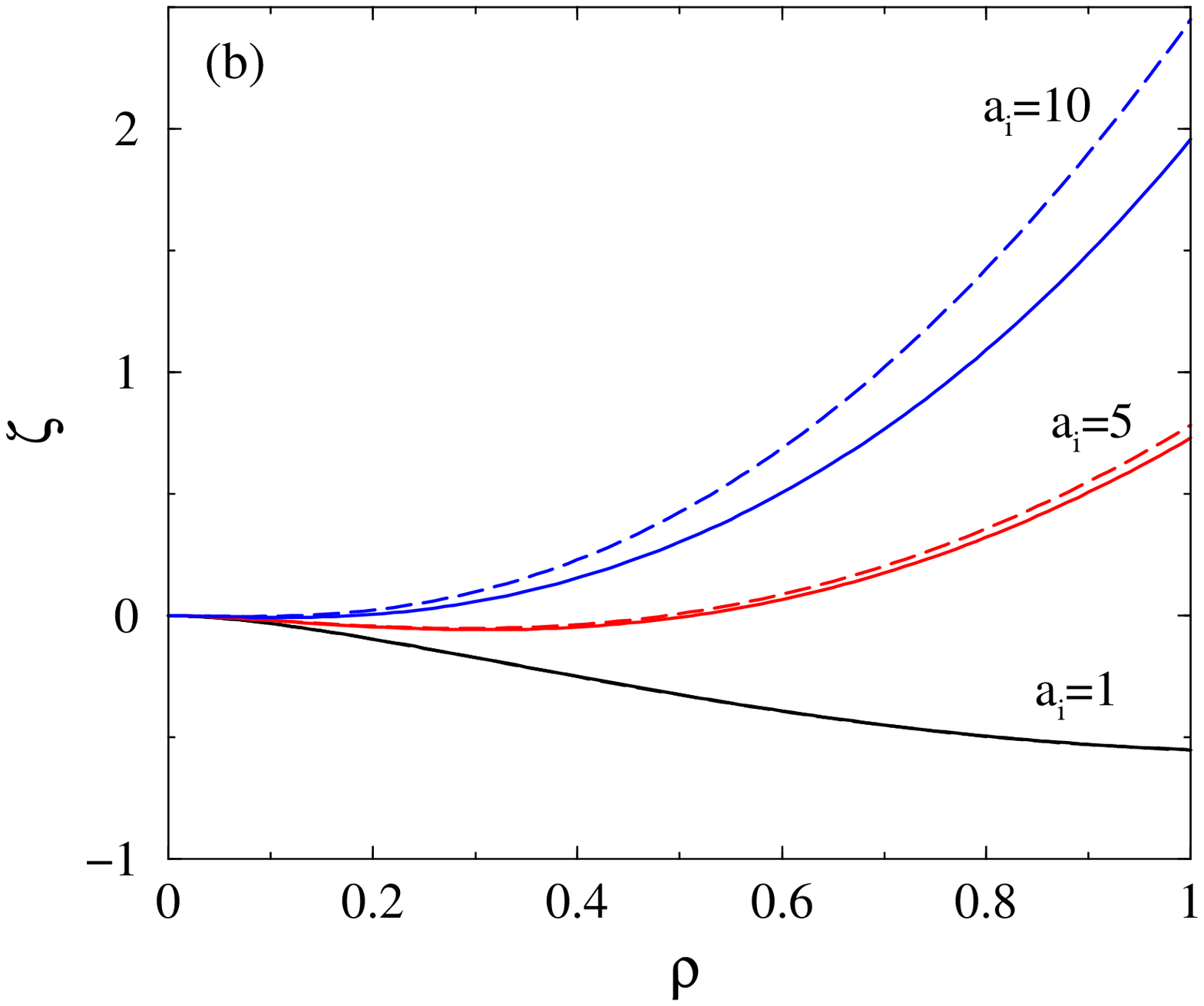}
}
\caption{Dimensionless displacement $\zeta(\rho)$ of the circular film illuminated by a 
circularly polarized light. Numerical (solid lines) and approximate (dashed lines) 
solutions for $\sigma=0.5$ and $f=1$ (a), $f=10$ (b), and different values of $a_{i}$.}
\label{fig:2}
\end{figure}

In order to verify the range of validity of the approximate solution (\ref{wrad}), 
full numerical simulations of Eqs.(\ref{fvkpol})-(\ref{bcTpol}) have been performed 
for the case of radial symmetry.
We used finite difference method solving the resulting set of nonlinear algebraic 
equations by Newton iterations.
The results of numerical calculations for the shape of irradiated film 
$\zeta(\rho)$ are shown in Fig.~\ref{fig:2} for the different values of the 
dimensionless gravity force $f$ and the parameter $a_i$ together with the 
approximate dependence (\ref{wrad}).
Even for large value of parameter $a_i$ the relative difference between numerical 
solution and the approximation (\ref{wrad}) does not exceed $25\%$ for $a_i=10$ and 
becomes smaller for smaller $a_i$.
For $a_i=1$ the numerical and approximate solutions are undistinguished in the 
Fig.~\ref{fig:2}.

%%% Figure 2 %%%
\subsection{Linearly polarized light}
\label{linear}
\begin{figure}
\resizebox{0.75\columnwidth}{!}{
\includegraphics{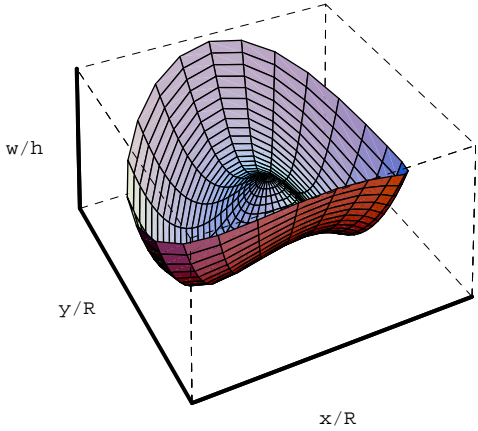}
}
\caption{The shape of the circular film illuminated by a linearly polarized light.}
\label{fig:3}
\end{figure}

Let us consider now the case of photoinduced film deformation caused by the linear 
polarized light (\ref{elw}), (\ref{ore}).
Expanding solutions of Eqs.(\ref{fvkpol})-(\ref{bcTpol}) in a Fourier series
\begin{eqnarray}
\label{fourier}
&& \zeta(\rho,\theta)=\sum_{n=0}^{\infty}\,\zeta_n(\rho)\,
\cos\left[ 2\,n\,(\psi-\theta) \right] \;,
\nonumber\\
&& \tau(\rho,\theta)=\sum_{n=0}^{\infty}\,\tau_n(\rho)\,
\cos\left[ 2\,n\,(\psi-\theta) \right] \;,
\end{eqnarray}
and assuming the weak light-film interaction $a_i<1$, $a_a<1$ [the condition 
$a_a<1$ allows to take into account only the zeroth and the first components in the 
Fourier series (\ref{fourier})] one finds
\begin{eqnarray}
\label{firsttwo}
\zeta(\rho,\theta) = \zeta_0(\rho)
- a_a\,\frac{\rho^2(3 - 2 \rho^2)}{6(3+\sigma)}\,\cos2(\psi-\theta) \;
\end{eqnarray}
for the vertical component of the displacement (see Fig.~\ref{fig:3}).
%

%%% Figure 3 %%%
\begin{figure}
\resizebox{0.75\columnwidth}{!}{
\includegraphics{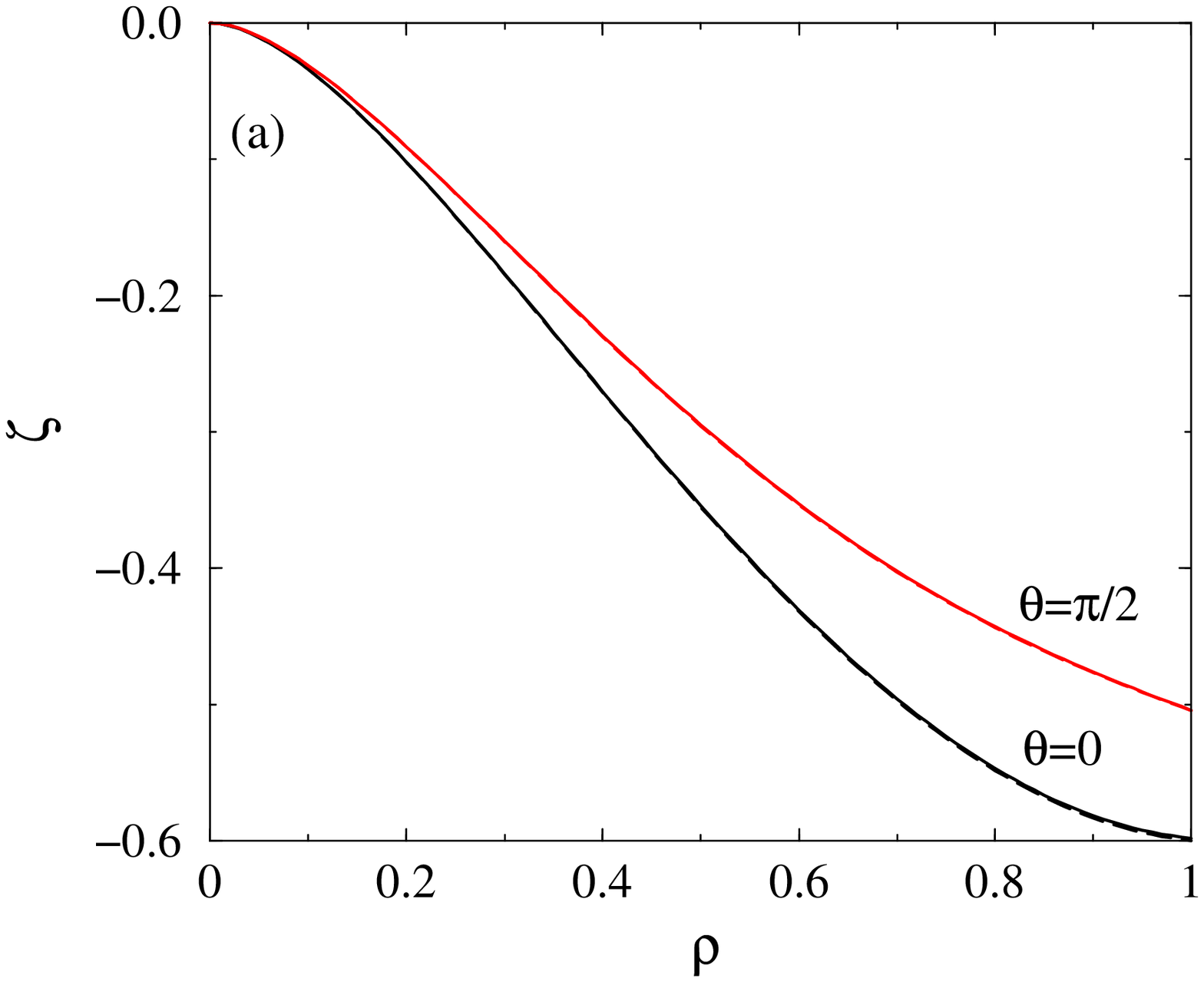}
}
\resizebox{0.75\columnwidth}{!}{
\includegraphics{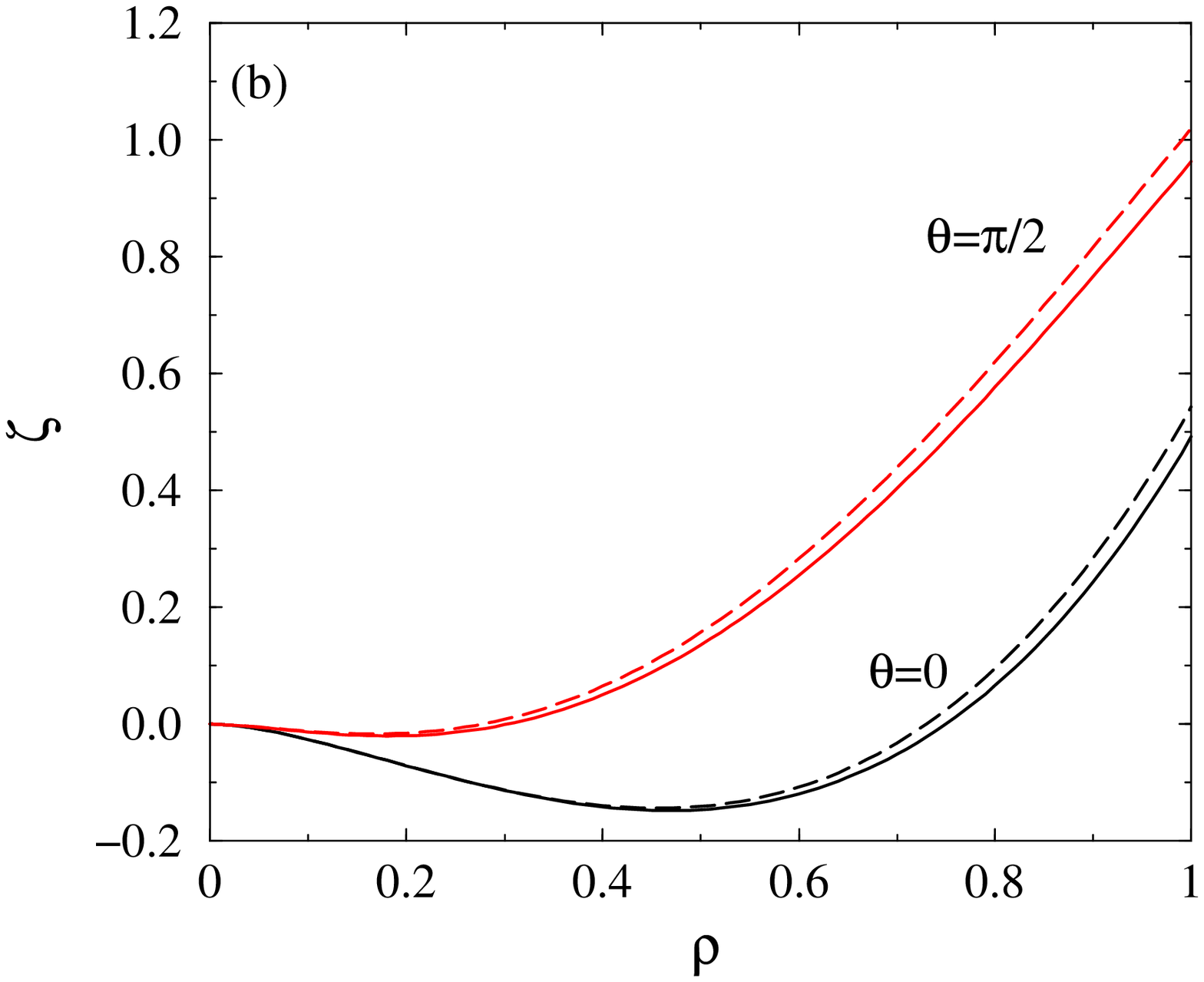}
}
\resizebox{0.75\columnwidth}{!}{
\includegraphics{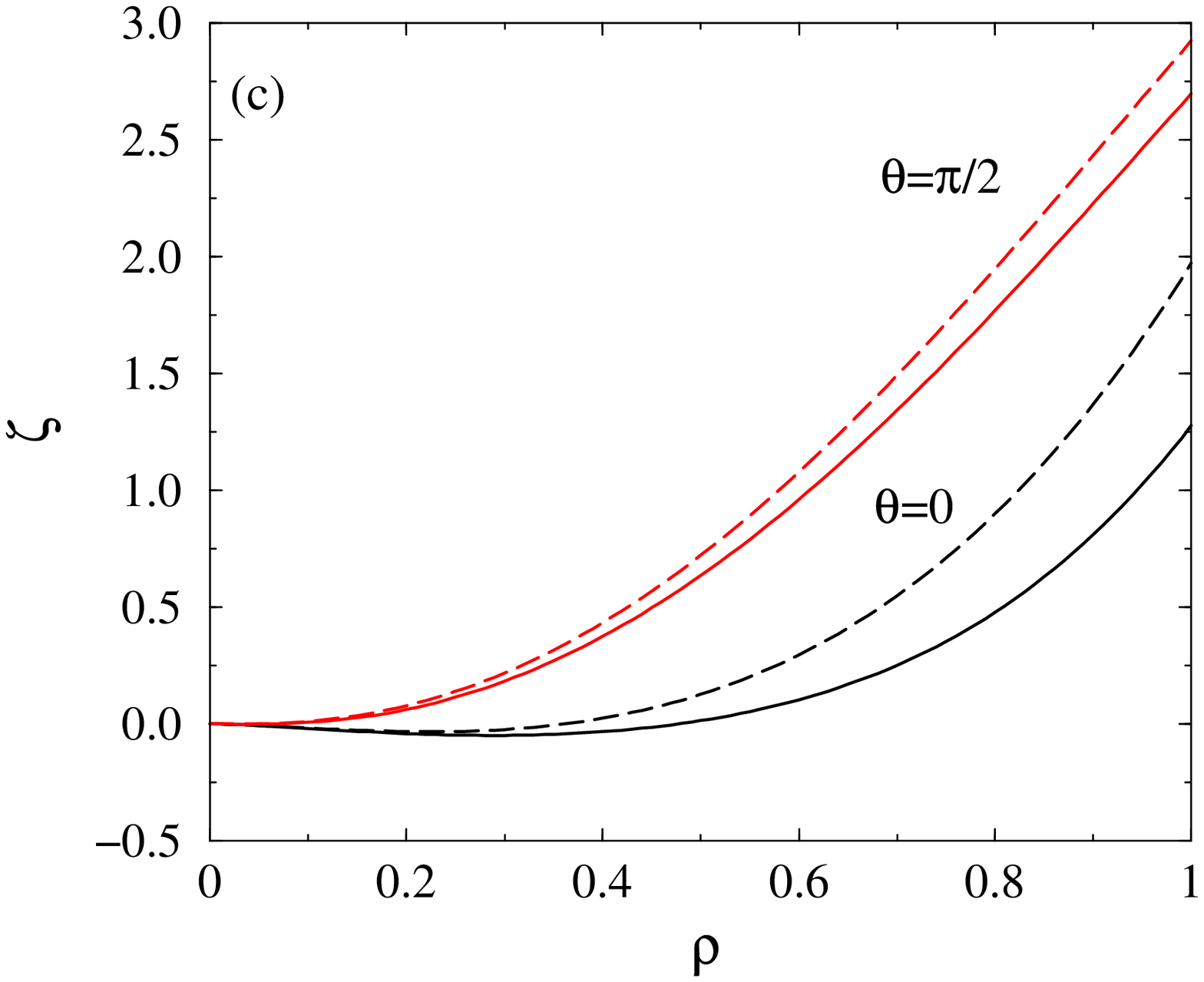}
}
\caption{Dimensionless displacement $\zeta(\rho)$ of the circular film illuminated 
by a linearly polarized light for $\psi=0$ in the cross sections $\theta=0$ and 
$\theta=\pi/2$. Numerical (solid lines) and approximate (dashed lines) solutions 
for $\sigma=0.5$ and $f=10$: $a_{i}=a_{a}=1$ (a), $a_{i}=a_{a}=5$ (b), and 
$a_{i}=a_{a}=10$ (c).}
\label{fig:4}
\end{figure}

For comparison full numerical simulations of Eqs.(\ref{fvkpol})-(\ref{bcTpol}) have 
been performed by use of finite difference method.
The results of calculations are shown in Fig.~\ref{fig:4}.
For small values of parameter $a_a$ the high order terms $\zeta_n$, $\tau_n$ in the 
expansion (\ref{fourier}) decrease for $n>1$ and the numerical solution for 
$\zeta(\rho,\theta)$ practically coincide with approximation (\ref{firsttwo}).
%

%%% Figure 4 %%%
\section{Bending of rectangular films}
\label{rectangular}
\begin{figure}
\resizebox{0.75\columnwidth}{!}{
\includegraphics{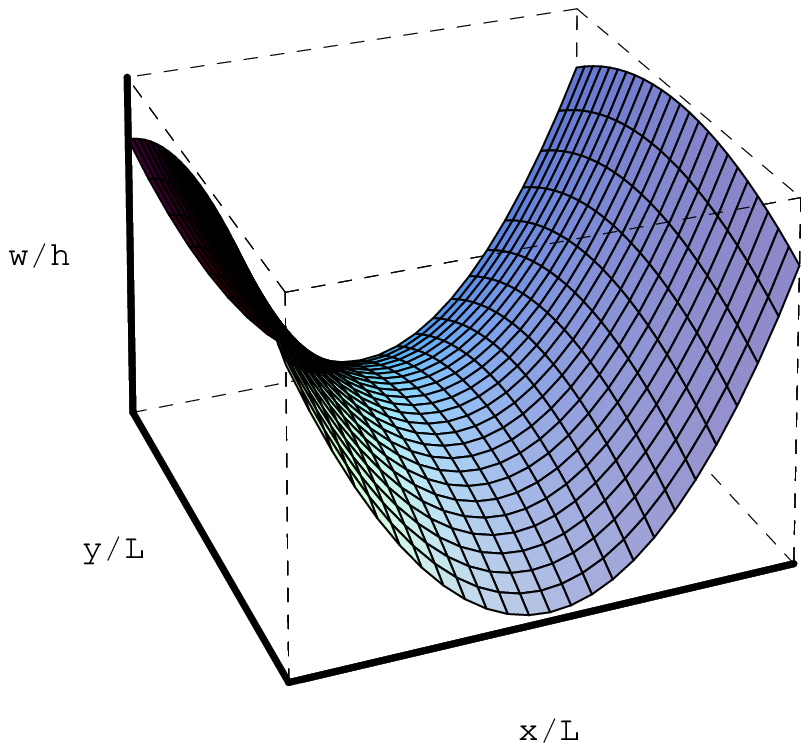}
}
\caption{The shape of the rectangular film with $A_a < A_i$ illuminated by a linearly 
polarized light for $\psi=0$.}
\label{fig:5}
\end{figure}

\begin{figure}
\resizebox{0.75\columnwidth}{!}{
\includegraphics{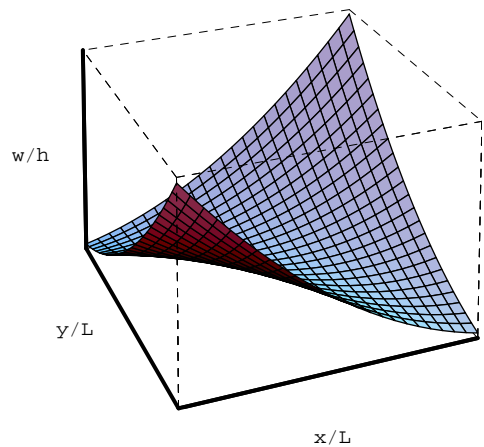}
}
\caption{The same as in Fig.~\ref{fig:5} for $\psi=\pi/4$.}
\label{fig:6}
\end{figure}

\begin{figure}
\resizebox{0.75\columnwidth}{!}{
\includegraphics{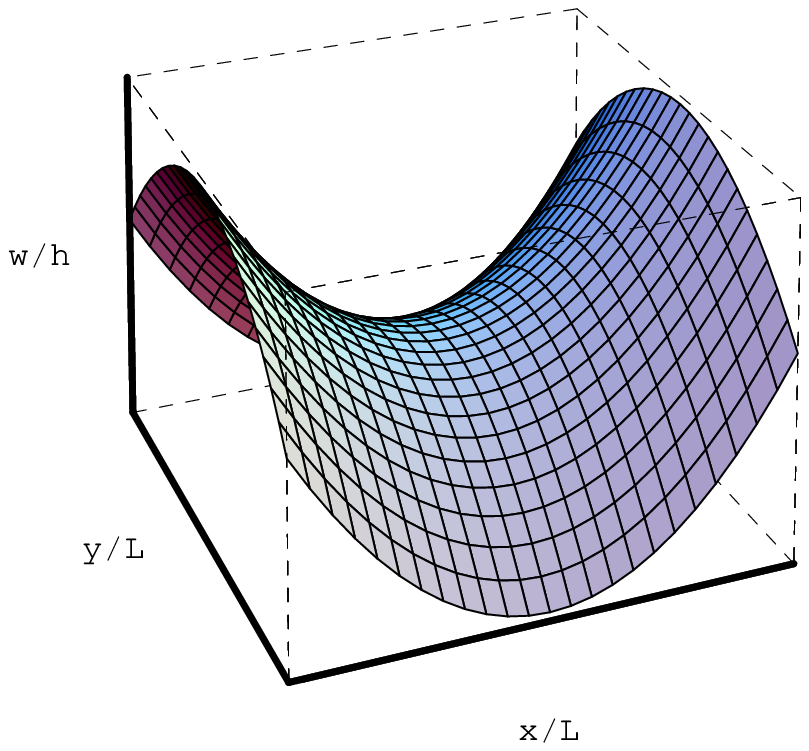}
}
\caption{The shape of the rectangular film with $A_a > A_i$ illuminated by a linearly 
polarized light for $\psi=0$.}
\label{fig:7}
\end{figure}

Let us now consider the case of rectangular film: $x \in (-L/2, L/2)$, 
$y \in (-L/2, L/2)$ interacting with a linearly polarized light.
We will consider the case of weak light-film interaction and use the linear theory 
(neglect the stretching contribution).
We will also neglect the action of the gravity force.
As a solution to Eq.(\ref{fvkl}) we choose
\begin{eqnarray}
\label{sol}
w = \frac{1}{2}a x^2+b x y+\frac{1}{2}c y^2 \;,
\end{eqnarray}
where $a$, $b$, and $c$ are some constants.
These constants can be found by using the boundary condition (\ref{bcdw}).
However a more simple way to find them is to introduce Eq.(\ref{sol}) into 
Eqs.(\ref{elbend}) and (\ref{bendint}) and get
\begin{eqnarray}
\label{elred}
\frac{1}{L^2} F =
\frac{D}{2}\left[ (a+c)^2 + 2(1-\sigma)(b^2-a c) \right]
\nonumber\\
-h^2 A_i (a+c) - h^2 A_a \left[ (a-c) \cos2\psi + 2 b
\sin2\psi \right] \;.
\end{eqnarray}
The function (\ref{elred}) has a minimum for
\begin{eqnarray}
\label{abc1}
&& a=h^2\,\frac{A_i(1-\sigma)+(1+\sigma)\,A_a\,\cos 2\psi}{D(1-\sigma^2)} \;,
\nonumber\\
&& c=h^2\,\frac{A_i(1-\sigma)-(1+\sigma)\,A_a\,\cos 2\psi}{D(1-\sigma^2)} \;,
\nonumber\\
&& b=h^2\,\frac{A_a\sin2\psi}{D(1-\sigma)} \;.
\end{eqnarray}
Thus under the action of linearly polarized light (\ref{elw}), (\ref{ore}) an 
initially flat molecular film takes the shape
\begin{eqnarray}
\label{shape}
w=\frac{1}{2}\,\kappa_1\,(x\,\cos\psi+y\sin\psi)^2
\nonumber\\
+\frac{1}{2}\,\kappa_2\,(-x\,\sin\psi+y\cos\psi)^2
\end{eqnarray}
which is characterized by the following two principal curvatures
\begin{eqnarray}
\label{curvat}
&& \kappa_1 = h^2\,\frac{A_i(1-\sigma) + A_a(1+\sigma)}{D\,(1-\sigma^2)} \;,
\nonumber\\
&& \kappa_2 = h^2\,\frac{A_i(1-\sigma) - A_a(1+\sigma)}{D\,(1-\sigma^2)} \;,
\end{eqnarray}
and has the equilibrium energy given by the expression
\begin{eqnarray}
F = -\frac{h^4 L^2}{D\,(1-\sigma^2)} \left[ A_i^2 (1-\sigma) +
A_a^2 (1+\sigma) \right] \;.
\end{eqnarray}
To characterize the global shape of the film it is convenient to introduce the mean 
curvature $H=(\kappa_1+\kappa_2)/2$ and the Gaussian curvature 
$K=\kappa_1\,\kappa_2$.
When $K > 0$ the point $(x=0,y=0)$ is an elliptic one and the film has a paraboloid 
shape.
For $K=0$ it becomes cylinder-like.
Let us assume first that $|\kappa_1| > |\kappa_2|$, then an irradiated film takes a 
shape close a cylindric one.
Note that this is probably the case in the experiments of 
Refs.~\cite{Ikeda_2003,Yu_2003} where figures show that polymer films are bent in a 
cylinder-like fashion.
In this case when the light is polarized along the $x$-axis ($\psi=0$) the bending 
occurs around the $y$-axis (see Fig.~\ref{fig:5}).
Under the action of light polarized along one of the film diagonals 
($\psi=\pi/4,3\pi/4$) the film bends around a diagonal (see Fig.~\ref{fig:6}).
This behavior is in a full agreement with the results of the 
Refs.~\cite{Ikeda_2003,Yu_2003}.
When $K < 0$ the film takes a saddle-like shape and the corresponding shape profile 
is shown in Fig.~(\ref{fig:7}).
Note that this probably the case in the experiments of 
Ref.~\cite{Camacho_Lopez_2004} where such of kind of deformation was observed for 
liquid-crystal elastomers with azo-dyes.
%

%%% Figure 5 %%%
%

%%% Figure 6 %%%
%

%%% Figure 7 %%%
\begin{figure}
\resizebox{0.75\columnwidth}{!}{
\includegraphics{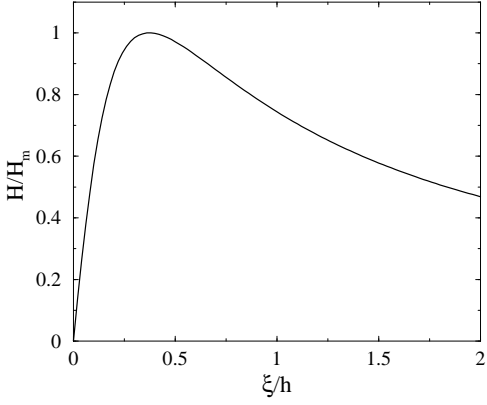}
}
\caption{The normalized mean curvature $H/H_{m}$ as function of scaled extinction 
length $\xi/h$.}
\label{fig:8}
\end{figure}

The principal curvatures of the film $\kappa_1$ and $\kappa_2$ are linearly 
proportional to the maximum population of the {\em cis}-isomers in the film 
${\cal N}_0$ and they are a non-monotonic function of the extinction length $\xi$.
In Fig.~\ref{fig:8} we presented the normalized mean curvature $H/H_m$ ($H_m$ is 
the maximum value of the mean curvature) as a function of $\xi$, restricted to 
$\xi<2 h$ ($h$ is the film thickness).
The last statement may be considered as a microscopical explanation of the 
conclusion which was drawn in Ref.~\cite{Warner_2004} in the frame of a 
phenomenological theory.
Note also that it is clear from Eqs.(\ref{AB}), (\ref{f}) and (\ref{curvat}) that 
for a decreasing number of {\em cis}-isomers in the film the curvature of the film 
becomes smaller and the film eventually returns to its initial shape.
This is happening in the experiments of Refs.~\cite{Ikeda_2003,Yu_2003} when the 
film is irradiated with light of the wavelength $> 540$~nm when the 
{\em cis}-isomeric state is depopulated.
%

%%% Figure 8 %%%
\section{Conclusion}
\label{concl}
In conclusion, we presented a nonlinear elastic theory which describes an 
anisotropic bending of molecular films under the absorption of polarized light.
Solving equations of equilibrium, we showed that a change in the polarization 
direction of light causes a corresponding change of the shape of the film.
We also showed that the curvature of an irradiated film is a non-monotonic function 
of the extinction coefficient.
\begin{acknowledgements}
Yu.~B.~G. acknowledges support from Deutsche Zentrum f{\"u}r Luft- und Raumfart 
e.V., Internationales B{\"u}ro des Bundesministeriums f{\"u}r Forschung und 
Technologie, Emil-Warburg-Stiftung, and Ministry of Education and Science of 
Ukraine, in the frame of a bilateral scientific cooperation between Ukraine and 
Germany, project No. UKR~05/055.
Financial support by the Deutsche Forschungsgemeinschaft Grant SFB~481 is 
gratefully acknowledged.
Partial support from National Academy of Sciences of Ukraine in the frame of 
``Fundamental properties of physical systems under extreme conditions'' is also 
acknowledged.
\end{acknowledgements}
\appendix
\section*{Appendix}
\label{appendA}
We will look for solutions of Eqs.(\ref{fvkpol})-(\ref{bcTpol}) in terms of  
Fourier series (\ref{fourier}).
Introducing Eqs.(\ref{fourier}) into Eqs.(\ref{fvkpol}), (\ref{ccpol}) and taking 
into account only the first two terms in the expansions (\ref{fourier}), we get
\begin{eqnarray}
\label{fvkr}
\Delta_\rho^2\,\zeta_0\,-\,\frac{1}{\rho}\,\frac{d}{d\,\rho}\,
\Big\{\frac{d\,\tau_0}{d\,\rho}\,\frac{d\,\zeta_0}{d\,\rho}
+\frac{1}{2}\frac{d\,\zeta_1}{d\,\rho}\,\frac{d\,\tau_1}{d\,\rho}
\nonumber\\
-2\,\frac{d}{d\,\rho}\left(\frac{1}{\rho}\,\tau_1\,\zeta_1\right)\Big\}
= -f \;,
\end{eqnarray}
\begin{eqnarray}
\label{ccr}
\Delta_\rho^2\,\tau_0\,+\,12 (1-\sigma^2) 
\frac{1}{2}\,\frac{1}{\rho}\,\frac{d}{d\,\rho}\,
\Big\{\left(\,\frac{d\,\zeta_0}{d\,\rho}\right)^2\,
\nonumber\\
+\,\frac{1}{2}\,\left(\,\frac{d\,\zeta_1}{d\,\rho}\right)^2
-2\,\frac{d}{d\,\rho}\left(\frac{1}{\rho}\,w^2_1\right)\,\Big\}
= 0 \;,
\end{eqnarray}
\begin{eqnarray}
\label{fvk1}
\left(\Delta_\rho - \frac{4}{\rho^2} \right)^2 \zeta_1 - \frac{1}{\rho}
\Big[ \frac{d}{d \rho} \left( \frac{d \zeta_0}{d \rho} \frac{d \tau_1}{d \rho} 
\right)
\nonumber\\
+\frac{d}{d \rho} \left( \frac{d \tau_0}{d \rho} \frac{d \zeta_1}{d \rho} \right)
-\frac{4}{\rho} \left( \zeta_1 \frac{d^2 \tau_0}{d \rho^2}
+\tau_1 \frac{d^2 \zeta_0}{d \rho^2} \right) \Big]
= 0 \;,
\end{eqnarray}
\begin{eqnarray}
\label{cc1}
\left(\Delta_\rho - \frac{4}{\rho^2} \right)^2 \tau_1 + 12(1-\sigma^2)\frac{1}{\rho}
\Big[ \frac{d}{d \rho} \left( \frac{d \zeta_0}{d \rho} \frac{d \zeta_1}{d \rho} 
\right)
\nonumber\\
-\frac{4}{\rho} \zeta_1 \frac{d^2 \zeta_0}{d \rho^2} \Big]
= 0 \;,
\end{eqnarray}
where 
$$\Delta_\rho=\frac{1}{\rho}\,\frac{d}{d\rho}\,\rho\frac{d}{d \,\rho}$$
is the radial part of the Laplace operator.
Inserting Eqs.(\ref{fourier}) into Eqs.(\ref{bcdwnpol})-(\ref{bcTpol}), we obtain 
the boundary conditions at $\rho=1$ for the zeroth harmonics in the form
\begin{eqnarray}
\label{bcr1}
\frac{d}{d \rho} \frac{1}{\rho} \frac{d}{d \rho}
\rho \frac{d \zeta_0}{d \rho} = 0 \;,
\end{eqnarray}
\begin{eqnarray}
\label{bcr2}
\frac{d^2\,\zeta_0}{d\,\rho^2}
+\frac{\sigma}{\rho}\,\frac{d\,\zeta_0}{d\,\rho} - a_i =0 \;,
\end{eqnarray}
\begin{eqnarray}
\label{bcr3}
\frac{d\,\tau_0}{d\,\rho} = 0 \;,
\end{eqnarray}
and for the first Fourier harmonics in the form
\begin{eqnarray}
\label{bc11}
\frac{d\,}{d\,\rho}\left(\Delta_\rho-\frac{4}{\rho^2}\right)\,\zeta_1
+4\,\frac{1-\sigma}{\rho^3}\,\left(\zeta_1-\rho\,\frac{d\,\zeta_1}{d\,\rho}\right)
\nonumber\\
-\frac{2}{\rho} a_a = 0 \;,
\end{eqnarray}
\begin{eqnarray}
\label{bc12}
\left(\Delta_\rho-\frac{4}{\rho^2}\right)\,\zeta_1
+\,\frac{1-\sigma}{\rho^2}\,\left(4\,\zeta_1-\rho\,\frac{d\,\zeta_1}{d\,\rho}\right)
- a_a = 0 \;,
\end{eqnarray}
\begin{eqnarray}
\label{bc13}
\tau_1 = 0 \;, \;\; \frac{d\,\tau_1}{d\,\rho} = 0 \;.
\end{eqnarray}
As it is seen from Eqs.(\ref{bc11}), (\ref{bc12}) the amplitude of the first 
harmonics is proportional to the light-film interaction parameter $a_a$.
Assuming that $a_a < 1$, we neglect all terms proportional $a_a^n$ with $n\geq 2$.
Under this assumption one can neglect the last two terms in Eqs.(\ref{fvkr}), 
(\ref{ccr}) and obtain
\begin{eqnarray}
\label{fvkra}
\Delta_\rho^2\,\zeta_0\,-\,\frac{1}{\rho}\,\frac{d}{d\,\rho}\,
\Big\{\frac{d\,\tau_0}{d\,\rho}\,\frac{d\,\zeta_0}{d\,\rho}\Big\} = -f \;,
\end{eqnarray}
\begin{eqnarray}
\label{ccra}
\Delta_\rho^2\,\tau_0\,+\, 
12(1-\sigma^2)\frac{1}{2}\,\frac{1}{\rho}\,\frac{d}{d\,\rho}\,
\Big\{\left(\,\frac{d\,\zeta_0}{d\,\rho}\right)^2\Big\} = 0 \;,
\end{eqnarray}
Integrating each equation, we get
\begin{eqnarray}
\label{fvkint}
\,\rho\,\frac{d}{d\,\rho}\frac{1}{\rho}\,\frac{d}{d\,\rho}\,
\rho\,\frac{d\,\zeta_0}{d\,\rho}\,
-\,\frac{d\,\tau_0}{d\,\rho}\,\frac{d\,\zeta_0}{d\,\rho}
=-\frac{1}{2}\,f\,(\rho^2-1),\nonumber\\
\,\rho\,\frac{d}{d\,\rho}\frac{1}{\rho}\,\frac{d}{d\,\rho}\,
\rho\,\frac{d\,\tau_0}{d\,\rho}\,
+\, 12(1-\sigma^2)\frac{1}{2}\,\left(\,\frac{d\,\zeta_0}{d\,\rho}\right)^2\, = 0 \;,
\end{eqnarray}
where the condition of regularity at the center $d\,\zeta_0/d\rho=0$ and the 
boundary conditions (\ref{bcr1}), (\ref{bcr3}) were used.
Equations (\ref{fvkint}) are simplified by introducing the new variables
\begin{eqnarray}
\label{newvar}
g = \frac{1}{\rho}\,\frac{d\,\zeta_0}{d\,\rho} \;, \;\;
\alpha = \frac{1}{4} \frac{1}{\rho}\,\frac{d\,\tau_0}{d\,\rho} \;, \;\;
z=\rho^2 \;.
\end{eqnarray}
Then Eqs.(\ref{fvkint}) become
\begin{eqnarray}
\label{fvkz}
&& \frac{d^2}{d\,z^2}\left(z\,g\right) - g\,\alpha\, =-\frac{f}{8}\,(1-\frac{1}{z}) 
\;,
\nonumber\\
&& \frac{d^2}{d\,z^2}\left(z\,\alpha\right) + \frac{3(1-\sigma^2)}{8}\,\,g^2 = 0 \;.
\end{eqnarray}
The boundary conditions (\ref{bcr1})-(\ref{bcr3}) at $z=1$ become
\begin{eqnarray}
\label{bcz}
&& \frac{d^2}{d\,z^2}\left(z\,g\right) = 0 \;,
\nonumber\\
&& 2\,z\,\frac{d\,g}{d\,z}+(1+\sigma)\,g - a_i = 0 \;,
\nonumber\\
&& \alpha=0 \;.
\end{eqnarray}
Assuming that $\epsilon \equiv 3(1-\sigma^2)/8 < 1$ is a small parameter we expand 
the functions $g(z)$ and $\alpha(z)$ into series
\begin{eqnarray}
\label{expans}
&& g=g_0+\epsilon\,g_1+\cdots \;,
\nonumber\\
&& \alpha=\alpha_0\,+\,\epsilon\,\alpha_1+\cdots \;.
\end{eqnarray}
Inserting (\ref{expans}) into Eqs.(\ref{fvkz}) we get
\begin{eqnarray}
\label{zero}
&\epsilon^0:\;&
\frac{d^2}{d\,z^2}\left(z\,g_0\right) - \,g_0\,\alpha_0\,
=-\frac{f}{8}\,(1-\frac{1}{z}) \;,
\nonumber\\
&\;&
\frac{d^2}{d\,z^2}\left(z\,\alpha_0\right) = 0 \;,
\end{eqnarray}
\begin{eqnarray}
\label{one}
&\epsilon^1:\;&
\frac{d^2}{d\,z^2}\left(z\,g_1\right)-\,g_0\,\alpha_1\, -\,g_1\,\alpha_0\, = 0 \;,
\nonumber\\
&\;&
\frac{d^2}{d\,z^2}\left(z\,\alpha_1\right)+\,g_0^2 = 0 \;.
\end{eqnarray}
In the same way the boundary conditions (\ref{bcz}) at $z=1$ can be expressed as
\begin{eqnarray}
\label{bcz0}
&\epsilon^0:\;&
\frac{d^2}{d\,z^2}\left(z\,g_0\right) = 0 \;,
\nonumber\\
&\;&
2\,z\,\frac{d\,g_0}{d\,z}+(1+\sigma)\,g_0 - a_i = 0 \;,
\nonumber\\
&\;&
\alpha_0 = 0 \;,
\end{eqnarray}
\begin{eqnarray}
\label{bcz1}
&\epsilon^1:\;&
\frac{d^2}{d\,z^2}\left(z\,g_1\right) = 0 \;,
\nonumber\\
&\;&
2\,z\,\frac{d\,g_1}{d\,z}+(1+\sigma)\,g_1 = 0 \;,
\nonumber\\
&\;&
\alpha_1 = 0 \;.
\end{eqnarray}
From Eqs.(\ref{zero}) and (\ref{bcz0}) one finds
\begin{eqnarray}
\label{g0}
&& g_0 = \frac{a_i}{1+\sigma} - \frac{f}{8} \left\{ \frac{1-\sigma}{2(1+\sigma)} + 
\frac{z}{2} - \ln z \right\} \;,
\nonumber\\
&& \alpha_0 = 0 \;.
\end{eqnarray}
%

% BibTeX users please use
% \bibliographystyle{}
% \bibliography{}

\begin{thebibliography}{10}
%

\bibitem{Bian_1998}
S.~Bian, L.~Li, J.~Kumar, D.Y.~Kim, J.~Williams, and S.K.~Tripathy,
Appl. Phys. Lett. {\bf 73}, 1817 (1998).

\bibitem{Kumar_1998}
J.~Kumar, L.~Li, X.L.~Liang, D.Y.~Kim, T.S.~Lee, and S.~Tripathy,
Appl. Phys. Lett. {\bf 72}, 2096 (1998).

\bibitem{Bublitz_2000}
D.~Bublitz, M.~Helgert, B.~Fleck, L.~Wenke, S.~Hvilsted, and P.S.~Ramanujam,
Appl. Phys. B {\bf 70}, 863 (2000).

\bibitem{Finkelmann_2001}
H.~Finkelmann, E.~Nishikawa, G.G.~Pereira, and M.~Warner,
Phys. Rev. Lett. {\bf 87}, 015501 (2001).

\bibitem{Ikeda_2003}
T.~Ikeda, M.~Nakano, Y.~Yu, O.~Tsutsumi, and A.~Kanazawa,
Adv. Mater. {\bf 15}, 201 (2003).

\bibitem{Yu_2003}
Y.~Yu, M.~Nakano, and T.~Ikeda,
Nature {\bf 425}, 145 (2003).

\bibitem{Camacho_Lopez_2004}
M.~Camacho-Lopez, H.~Finkelmann, P.~Palffy-Muhoray, and M.~Shelley,
Nature Materials {\bf 3}, 307 (2004).

\bibitem{Lendlein_2005}
A.~Lendlein, H.~Jiang, O.~J{\"u}nger, and R.~Langer,
Nature {\bf 434}, 879 (2005).

\bibitem{DeGennes_1975}
P.G.~De~Gennes,
C. R. Acad. Sci. Ser. B {\bf 281}, 101 (1975).

\bibitem{Warner_2003}
M.~Warner and E.M.~Terentjev,
{\em Liquid Crystal Elastomers} (Clarendon Press, Oxford, 2003).

\bibitem{Warner_2004}
M.~Warner and L.~Mahadevan,
Phys. Rev. Lett. {\bf 92}, 134302 (2004).

\bibitem{Gaididei_2002}
Y.B.~Gaididei, P.L.~Christiansen, and P.S.~Ramanujam,
Appl. Phys. B {\bf 74}, 139 (2002).

\bibitem{Pedersen_1998}
T.G.~Pedersen, P.S.~Ramanujam, P.M.~Johansen, and S.~Hvilsted,
J. Opt. Soc. Am. B {\bf 15}, 2721 (1998).

\bibitem{Corbett_2006}
D.~Corbett and M.~Warner,
Phys. Rev. Lett. {\bf 96}, 237802 (2006).

\bibitem{Landau_1986}
L.D.~Landau and E.M.~Lifshitz,
{\em Theory of Elasticity} (Pergamon, Oxford, 1986).

\bibitem{Stoker_1983}
J.J.~Stoker,
{\em Nonlinear Elasticity} (Gordon and Breach, New York, 1983).

%
\end{thebibliography}
%
% Non-BibTeX users please use

%
\end{document}